\documentclass[aps,prl,twocolumn,showpacs,floatfix,superscriptaddress,groupedaddress]{revtex4-1}  
\usepackage{graphicx}  
\usepackage{dcolumn}   
\usepackage{bm}        
\usepackage{amssymb}   
\usepackage{gensymb}
\usepackage{color}  
\renewcommand{\vec}[1]{\boldsymbol{\mathbf{#1}}}
\usepackage{hyperref}
\usepackage{xcolor}

\hypersetup{
    colorlinks,
    linkcolor={blue!50!black},
    citecolor={blue!50!black},
    urlcolor={blue!80!black}
}
\begin{document}

\widetext

\title{Imaging Dirac Fermions Flow Through a Circular Veselago Lens}

\author{B. Brun$^{1}$, N. Moreau$^1$, S. Somanchi$^2$, V.-H. Nguyen$^1$,  K. Watanabe$^3$, T. Taniguchi$^3$, J.-C. Charlier$^1$, C. Stampfer$^2$ \& B. Hackens$^1$}

\affiliation{
$^1$IMCN/NAPS, Universit\'e catholique de Louvain (UCLouvain), B-1348 Louvain-la-Neuve, Belgium \\
$^2$ JARA-FIT and 2nd Institute of Physics - RWTH Aachen, Germany \\
$^3$ National Institute for Materials Science, Namiki, Tsukuba, Japan
}

\date{\today}

\begin{abstract}

Graphene charge carriers behave as relativistic massless fermions, thereby exhibiting a variety of counter-intuitive behaviors. In particular, at p-n junctions, they behave
as photons encountering a negative index media, therefore experiencing a peculiar refraction known as Veselago lensing. 
However, the way Dirac fermions flow through a Veselago lens remains largely unexplored experimentally.
Here, a novel approach to create a movable and tunable circular p-n junction in graphene is proposed, using the polarized tip of a scanning gate microscope.
Scanning the tip in the vicinity of a graphene constriction while recording the device conductance yields images related to the electron flow through a circular Veselago lens, 
revealing a high current density in the lens core, as well as two low current density zones along transport axis. 
Tight-binding simulations reveal the crucial role of the p-n junction smoothness on these phenomena. 
The present research adds new dimensions in the control and understanding of Dirac fermions optical elements,
a prerequisite to engineer relativistic electron optics devices.
\end{abstract}

\pacs{}
\maketitle


Optic-like manipulation of electron beams could in principle be achieved in p-n junctions, provided that the junction transmission is near perfect. 
However, in most cases, in particular in semiconductors, conventional p-n junctions are poor electron transmitters due to the inherent depletion of carriers in the vicinity of the junction. 
The discovery of graphene \cite{Novoselov-2005} changed this paradigm. Since charge carriers in graphene behave as massless Dirac fermions, Klein-like tunneling \cite{Klein-1929, Young-2009} 
is at play, allowing perfect transmission for charge carriers impinging on a p-n junction perpendicularly. In addition, a diverging flow of Dirac fermions is refocused at a p-n interface, 
an effect known as Veselago lensing \cite{Veselago-1968}.

Using this negative refraction, invisibility cloak has been realized for microwave photons \cite{Schurig-2006}, and perfect electromagnetic lenses have been proposed \cite{Pendry-2000}.
The possibility of forming such Veselago lenses using p-n junctions in graphene \cite{Cheianov-2007, Milovanovic2015} as well as the resulting 
caustics \cite{Cserti-2007,Mu-2011,Agrawal-2014, Wu-2014, Logermann-2015, Lu-2018,Zhang-2018} have attracted a lot of attention from a theoretical point of view. 
Recent works suggested that Veselago lensing could be used to create highly focused electron beams \cite{Ming-Hao-2017}, and even a two-dimensional scanning
Dirac fermions microscope \cite{Boggild-2017}. But the experimental challenges required to materialize these visionary ideas have not yet been overcome. 
Though few experiments reported signatures of Veselago lensing \cite{Lee-2015, Chen-2016}, the lack of tunability of such devices kept these discoveries to 
the state of ``proof of concept''. The use of a scanning tunneling microscopy (STM) tip to induce circular p-n junctions and probe the resulting local density of 
states unveiled beautiful phenomena such as whispering gallery modes \cite{Zhao-2015, Jiang-2017} and quasi-bound states \cite{Lee-2016}. 
However, though these approaches reveal the rich internal electronic structure of p-n nano-islands, they cannot probe the current density \textit{through}
these p-n junctions and in their vicinity, an essential parameter to control when designing Veselago-lensing-based devices.

In the present letter, a novel way to image current density through a tunable circular p-n junction, using a scanning gate microscope, is proposed. 
Scanning gate microscopy (SGM) consists in scanning a polarized metallic tip, acting as a local gate above a device's surface, and mapping out tip-induced conductance changes \cite{Eriksson-1996}. 
Initially developed to investigate transport in III-V semiconductor heterostructures \cite{Topinka-2001,Jura-2009, Kozikov-2013}, it has been extended by several groups 
to investigate transport in graphene devices \cite{Schnez-2010,Pascher-2012,Garcia-2013,Cabosart-2017,Bhandari-2016,Dou-2018,Shaohua-2016}.
However, very few SGM experiments have been conducted yet on clean encapsulated graphene devices. Hence, it has proven difficult to observe 
electron-optics behavior in graphene using SGM, despite several theoretical predictions \cite{Mrenca-2015,Mrenca-2016,Mrenca-2017}.

Here, the polarized SGM tip is used to create a movable and tunable circular Veselago lens. Applying a large voltage on the conductive tip induces circular p-n junction, 
whose potential can be controlled in amplitude and decay length.
The tip scans in the vicinity of a constriction etched in a high mobility encapsulated graphene flake.
Mapping the device conductance as a function of tip position yields images related to the electron flow through this Veselago lens.
Surprisingly, transmission efficiency is drastically reduced when placing the lens next to the constriction entrances, and significantly enhanced when placing the lens at the constriction center. 
Tight-binding simulations indicate that SGM images partly reflect the current density distribution around the Veselago lens.
The present work also emphasizes the role of the p-n interfaces sharpness and opens a door towards high precision optical elements design for Dirac fermions optics.

\begin{figure}
\includegraphics[width = 1.0 \linewidth]{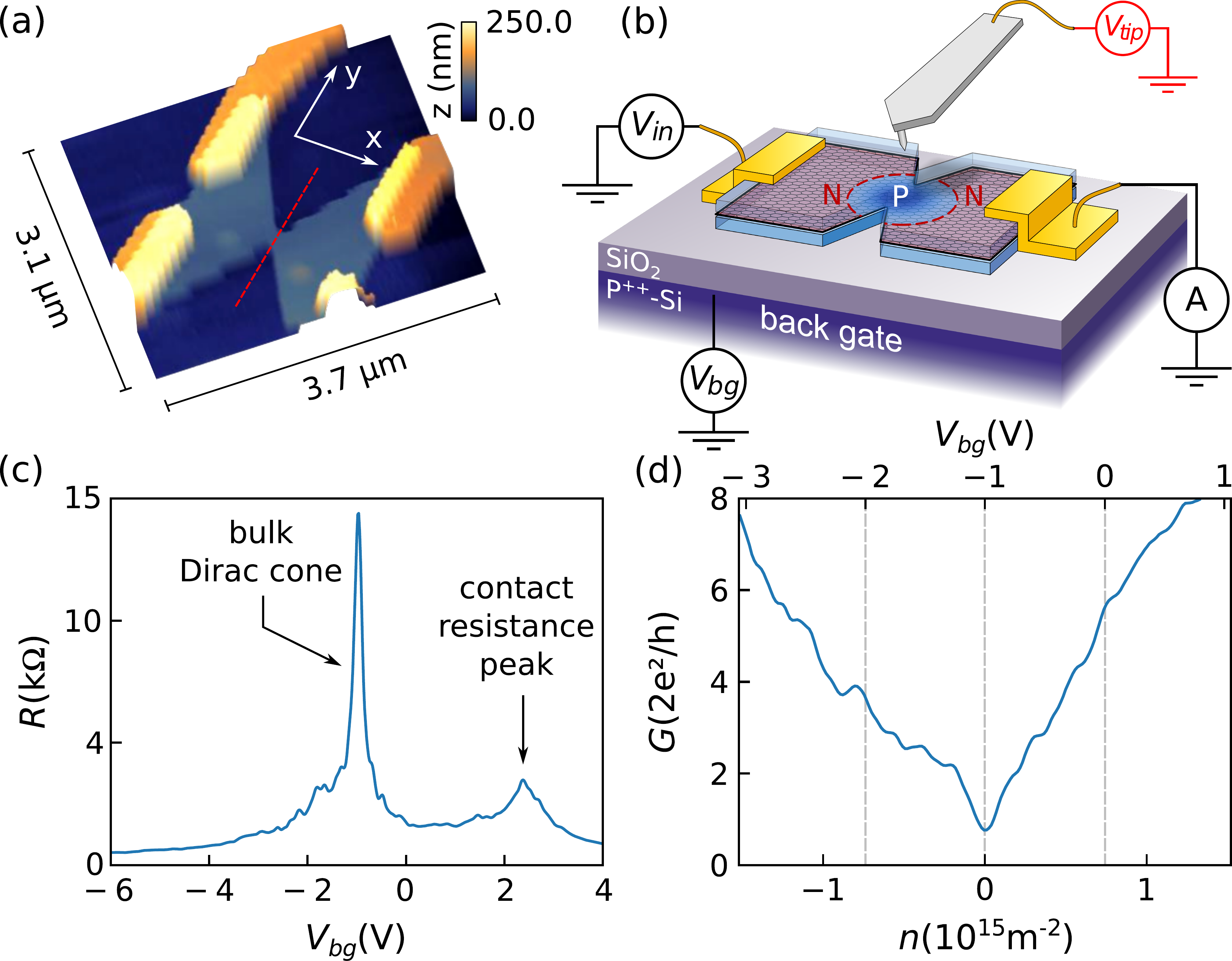}
\caption{\label{fig1} \textbf{Graphene constriction:}
(a) AFM image of the studied device (b) Scheme of the experiment: the charged AFM tip locally creates a n-p-n junction in the graphene flake.
(c) Resistance as a function of backgate-voltage $V_{bg}$ recorded in 2 contacts by standard lock-in technique at 4K, using an AC excitation of 35 $\mu$V at 77 Hz. 
The main Dirac cone is at a gate voltage of $V_{bg}$ = -1 V, and a smaller Dirac cone around $V_{bg}$ = +2.5 V is due to one of the ohmic contacts. (d) Conductance
as a function of $V_{bg}$ around the main Dirac cone.}
\end{figure}

The sample consists of a single layer graphene flake encapsulated between two 20 nm-thick h-BN flakes, deposited on top of a doped Si substrate covered by a 300 nm-thick SiO$_2$ layer.
A 250 nm-wide constriction is defined by etching \cite{Terres-2016} (see Fig. \ref{fig1}a). The mean free path is of the order of the contacts separation (see supplemental materials \cite{Supplemental}).

The sample is thermally anchored to the mixing chamber of a dilution refrigerator in front of a cryogenic scanning probe microscope \cite{Hackens-2010}.
Most of the data presented here are recorded  at a temperature of 4~K, except Fig.~\ref{fig3}(d-f) recorded at a base temperature of 28~mK.
The conductance $G$ of the device is measured between two line-contacts (as shown in Fig.~\ref{fig1}b), and contact series resistances of 2 k$\Omega$ are taken into account.
One of the contacts' resistance exhibits a strong dependence on back-gate voltage $V_{bg}$, and presents a secondary resistance peak around $V_{bg} = + 2.5$ V (Fig. \ref{fig1}c).
Note that this does not play any role here since the absolute $G$ value is not discussed, only its relative variations.

\begin{figure}
\includegraphics[width = 1.0 \linewidth]{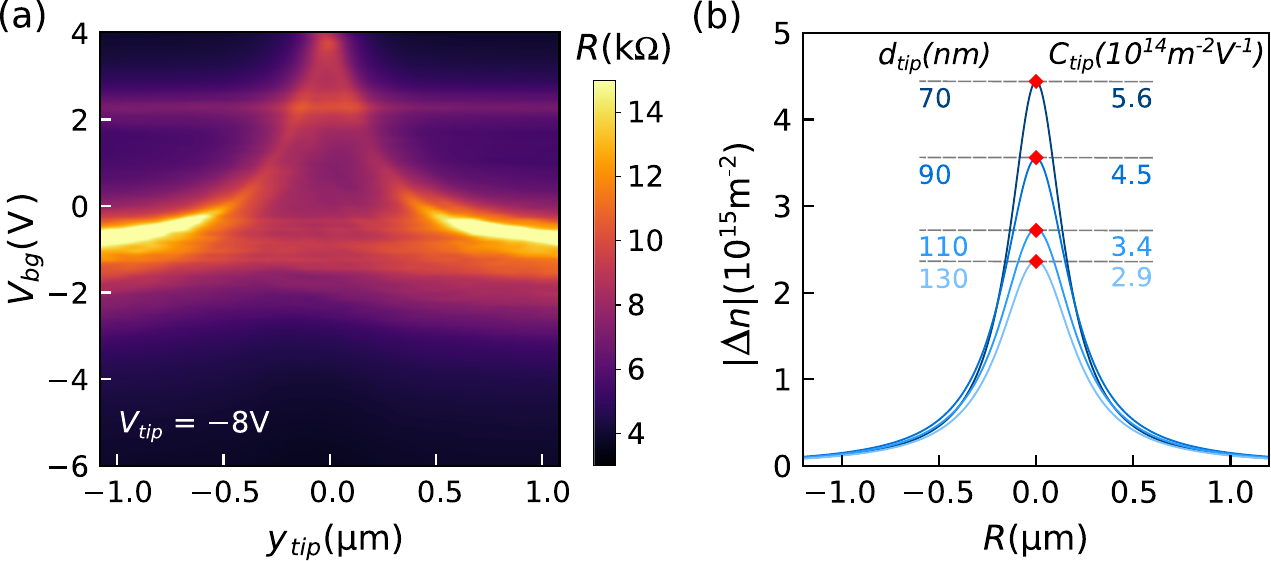}
\caption{\label{fig2} \textbf{Evaluation of the tip-induced potential:}
(a) Resistance as a function of tip position along red dashed line Fig. \ref{fig1}a (zero being the center of the constriction), and $V_{bg}$. $d_{tip} = 90$ nm, $V_{tip} = -8$ V. (b) Tip lever-arm parameter evolution with $d_{tip}$, for $V_{tip} = -8$ V }
\end{figure}

\begin{figure}
\includegraphics[width = 1.0 \linewidth]{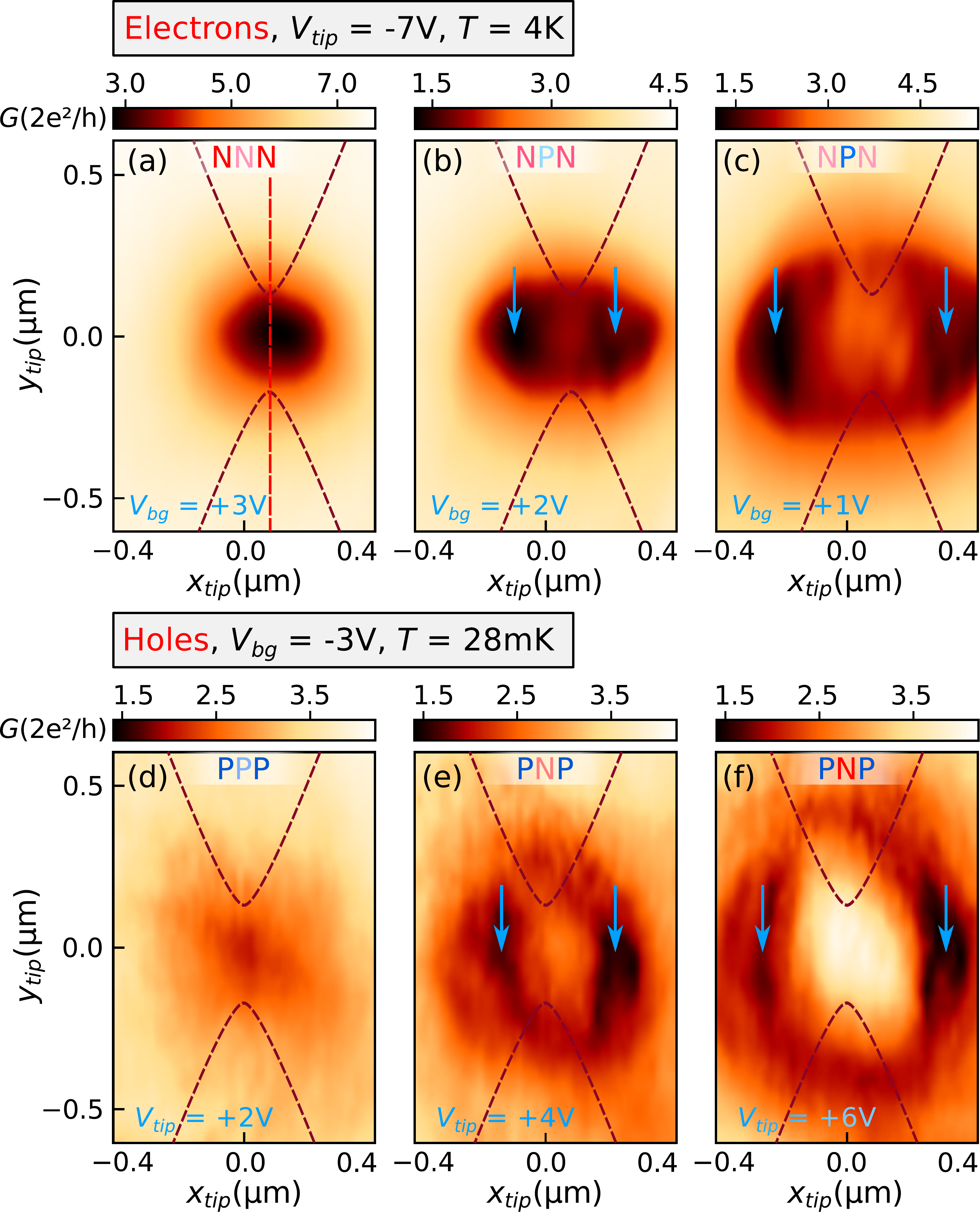}
\caption{\label{fig3}\textbf{SGM images evolution with carriers density and temperature:} (a) to (c) SGM images on electron side at 4K, for $V_{tip}$ = -7 V and
$d_{tip} = 120$ nm, inducing a density change $\Delta  n_{max} \sim -2.3\times10^{15} $ m$^{-2}$. (a) Bulk density $n = 3\times 10^{15} $ m$^{-2}$;
(b) $2.2\times 10^{15} $m$^{-2}$; (c) $1,5\times 10^{15}$ m$^{-2}$. (d) to (f): SGM images at 28 mK on holes side for $n = -1.5\times 10^{15} $ m$^{-2}$ and $d_{tip} = 120$ nm for
different $V_{tip}$ corresponding to $\Delta n_{max}$ = (d) $1.2\times 10^{15} $ m$^{-2}$; (e) $2.5\times 10^{15} $m$^{-2}$; (f) $3.7\times 10^{15} $ m$^{-2}$.
}
\end{figure}

We now discuss transport under the influence of the scanning tip. The biased tip locally changes the carriers density $n$, in an isotropic way, leading to a Lorentzian evolution of $n$, centered at the tip position. 
When a p-n junction is induced by the biased tip, this Lorentzian profile sharpness has a crucial influence on Dirac fermion semi-classical trajectories in its vicinity \cite{Allain-2011}. 
It is therefore important to extract this profile from our experimental data. 
For this, we scan the tip along a line perpendicular to the constriction transport axis 
(red dashed line in Fig.~\ref{fig1}a and \ref{fig3}a), at a distance $d_{tip}$ from the sample surface, and map the device resistance $R$ as a function of back-gate voltage (Fig. \ref{fig2}a). 
The \emph{locus} of maxima in the line profile mapping as a function of $V_{bg}$ draws a Lorentzian shape centered at the constriction.
This \emph{locus} corresponds to the tip-position-dependent value of $V_{bg}$ required to locally reach the Dirac point at the center of the constriction, 
hence the change in $n$ required to compensate the tip-induced local change of density. 
In other words, the Lorentzian shape in Fig. \ref{fig2}a directly maps the effect of the tip-induced potential on the local carrier density, 
provided that the $V_{bg}$-axis is properly scaled into the local carrier density change $\Delta n$.
The Lorentzian maximum linearly scales with tip voltage $V_{tip}$, allowing its characterization by a simple lever-arm parameter that depends 
on tip-to-graphene distance $d_{tip}$, as shown Fig. \ref{fig2}b.
The half width at half maximum $R_{tip}$ depends also on $V_{tip}$ and $d_{tip}$ and varies between 200 and 300 nm (see supplemental materials).

With this fully-characterized tip influence, the SGM images (i.e. conductance as a function of tip position) can be analyzed. They are obtained by scanning the tip above the sample at fixed $V_{tip}$ and $V_{bg}$.
Figure \ref{fig3} shows such SGM maps, in different conditions of bulk carrier densities, tip voltages and temperature $T$.
Figure \ref{fig3}a-c shows the evolution of the SGM contrast with carrier density. $V_{bg}$ is chosen such that bulk transport is governed by electrons.
In Fig. \ref{fig3}a, the highest tip-induced density change ($\Delta  n_{max} \sim -2.3\times10^{15}$ m$^{-2}$) is lower than the bulk electron density 
($3\times10^{15}$ m$^{-2}$). This therefore corresponds to a n-n'-n configuration, inducing a quasi-isotropic decrease of $G$ as the negatively-biased tip moves away from
the constriction center. This means that $G$ is governed by the constriction region, and that reducing the carriers density in the center of the constriction
reduces the total transmission, as naturally expected. This prominent influence of the constriction on transport justifies the validity of the above-described method to characterize
the tip-induced potential.
For lower bulk densities (Fig. \ref{fig3}b-c), the tip bias creates a circular n-p-n junction.
In this case, an important qualitative change is observed. Contrary to Fig. \ref{fig3}a, we observe a local maximum of $G$ when the tip is placed at the very center of the constriction, 
surrounded by an elliptic region of reduced conductance. 
Interestingly, $G$ change is not isotropic anymore, and two spots of lower $G$ are clearly visible on both sides of the constriction, as indicated by arrows in Fig. \ref{fig3}b-c. 

A similar behavior is observed for hole-type carriers, as shown in Fig. \ref{fig3}d-f. Here, the bulk hole density is fixed ($V_{bg} = -3$ V) and $V_{tip}$ is varied.
Once again, reducing the hole density below the tip leads to a quasi-isotropic single spot of reduced $G$ (Fig. \ref{fig3}d). But if $V_{tip}$ is chosen such that it
locally changes the charge carriers sign (i.e. a circular p-n-p region is created), a local maximum of $G$ is observed when placing the tip at the very center of the constriction,
and two regions of lower $G$ are visible when placing the tip on both sides of the constriction along transport axis (Fig. \ref{fig3}e-f). 
Note that data in Fig. \ref{fig3}d-f are recorded at $T = 28$ mK, which does not qualitatively change the main observations compared to Fig. \ref{fig3}a-c, recorded at 4 K. 
Noteworthy, the main features are really robust to temperature changes, and still observable up to 100 K (see supplemental materials).

\begin{figure}[h]
\includegraphics[width = 1.0 \linewidth]{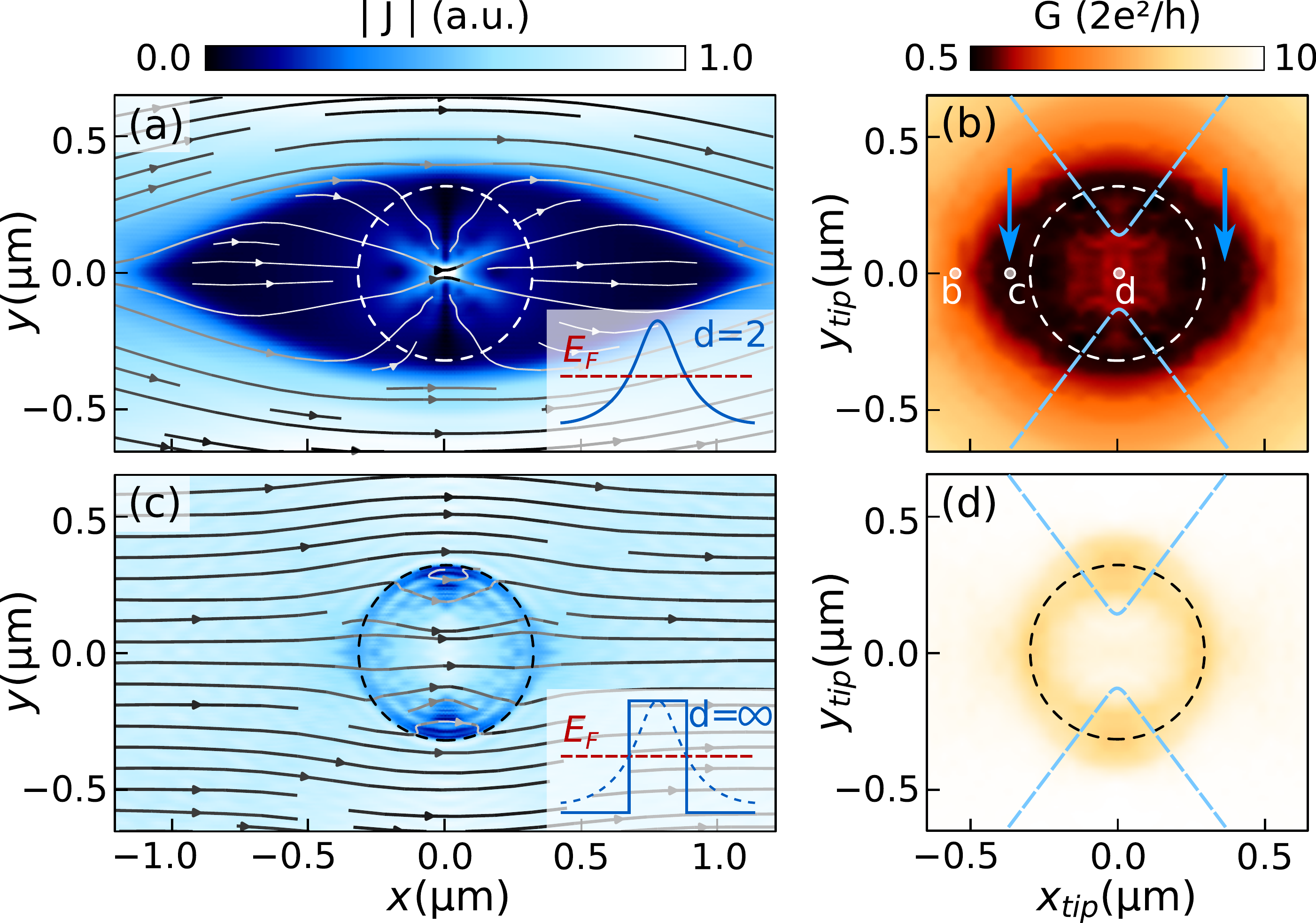}
\caption{\label{fig4} \textbf{Tight-binding simulations:}
Simulations of current density $\vec{J}$ in a graphene sheet around the fixed tip potential of Eq. (\ref{Lorentzian}) characterized by the parameters $V_m$, $E$ and $d$ (left image in each couple),
and simulated SGM maps obtained by displacing the same potential around a 250 nm-wide constriction (right image in the couple). 
(a-b) n-p-n configuration with $d =2$ and $V_m$ = $2E$.
(c-d) Abrupt circular p-n junction with $d =\infty$ and $V_m$ = $2E$.
The chosen energy $E$ corresponds in all cases to $n = 1.5\times 10^{15} \rm m^{-2}$, and $\Delta n_{max} = -3\times 10^{15} \rm m^{-2}$.
Dashed circle indicates the \emph{locus} corresponding to zero charge density. Insets: schematics of potential profile vs position.
}
\end{figure}

In the following, we discuss the origin of the observed features, and evidence a correspondence between the SGM maps 
and the current density pattern inside and around a tip-induced circular Veselago lens.
It should first be stressed that a model of the experiment assuming diffusive transport in the constriction region cannot 
reproduce our observations (see section V of the supplemental materials \cite{Supplemental}).
The measured mobility is limited by the ohmic contacts separation, indicating ballistic transport on a scale larger than the radius of the circular p-n junction defined by the tip.
This is corroborated by the observation of Fabry-P\'erot interference in the n-p-n and p-n-p configurations \cite{Fabry-Perot}.

To understand the contrast in SGM images, tight-binding simulations of transport through model graphene devices are performed, using a home-made recursive Green function code \cite{Nguyen-2010}.
The charge carrier flow around a tip-induced potential is first investigated in an infinite graphene sheet without the constriction, as shown in Fig. \ref{fig4} a and c.
The tip-induced perturbation is included by imposing a potential
\begin{equation}\label{Lorentzian}
V(\vec{r}) = \frac{V_m}{1+ \left(\frac{\Arrowvert \vec{r}-\vec{r}_{tip} \Arrowvert}{R_{tip}}\right)^d} 
\end{equation}
where $\vec{r}_{tip}$ is the tip position, $R_{tip}$ the potential decay length and $d$ the decay exponent.
The current density $\arrowvert \vec{J} \arrowvert$ is computed at a given energy $E$. Note that all distances, energies and potential are scaled to match the experimental parameters.
Figure \ref{fig4}a presents the current density around a Lorentzian potential (d=2 in Eq.(\ref{Lorentzian})) fixed at $\vec{r}_{tip} = (0,0)$, for $R_{tip}= 300$ nm and $V_{m}$ = $2E$, 
therefore inducing a circular p-n junction. In this configuration, $\arrowvert \vec{J} \arrowvert$ is maximal at the center of the junction.
Figure \ref{fig4}b shows the computed SGM map when the perturbation is scanned around a 250 nm-wide constriction (represented by blue dashed lines), i.e. the system transmission as a 
function of the Lorentzian potential position.

These tight-binding simulations capture the essential features observed in the experimental SGM images (Fig. \ref{fig3}), and the comparison with current density maps shines light 
on the mechanisms at play in the experiment.
Indeed, scanning the tip in the vicinity of a constriction yields SGM images that are qualitatively similar to the current density pattern around and through the tip-induced circular p-n junction. 
This result contrasts with SGM images obtained on GaAs quantum point contacts, which take advantage of the tip-induced backscattering to image the unperturbed electron 
wavefunctions \cite{Topinka-2000}, or trajectories through the bulk disordered potential \cite{Topinka-2001, Braem-2018}. 
In graphene, backscattering is prevented by pseudo-spin conservation \cite{Allain-2011}, leading to Klein tunneling, and this reverses the paradigm of SGM experiment.
In turn, this allows to study the Dirac fermion flow through the tip potential.

Klein tunneling is characterized by a perfect transmission for trajectories impinging perpendicularly to a p-n junction. 
Beside this effects, trajectories crossing the junction at finite angles are refracted with a negative index.
Given the circular geometry in our experiment, which offers a large collection of incidence angle, these two manifestations of the same physical 
phenomenon do contribute to the high $\arrowvert \vec{J} \arrowvert$ spot at the p-n junction center. 

The smoothness of the p-n interface strongly impacts the transmission probability at finite incident angle \cite{Cheianov-2006, Allain-2011, Chen-2016, Libisch2017}.
For equal carrier densities in the p and n regions, this probability for an incidence angle $\phi$ is given by $ T(\phi) = e^{-\pi k_F \omega sin^2(\phi)}$, 
where $k_F$ is the Fermi wave-vector, and $\omega$ the junction width.
This indicates that trajectories impinging with an incident angle up to $\sim 20\degree $ are refracted with a negative index and contribute to the total transmission in our experimental conditions.

The SGM images reveal two main features of the charge flow through the p-n junction:
(i) The local maximum of transmission, observed when the tip is placed at the center of the constriction, reflects the high current density at the lens core in the n-p-n or p-n-p configuration. 
(ii) The reduced conductance spots observed when the tip is placed slightly away from the constriction, indicated by arrows in Fig. \ref{fig4}b and very similar to 
the experimentally observed spots in Fig. \ref{fig3}, are the \emph{loci} of all the tip positions for which the transmission of Dirac fermions through the constriction is strongly reduced. 
This is reminiscent of the effect visible in Fig. \ref{fig4}a where the tip also induces low current density (dark) regions on the sides of the tip-induced p-n junction. 

Interestingly, the smoothness of the p-n junctions plays a crucial role in the emergence of the low conductance and low $\arrowvert \vec{J} \arrowvert$ points. 
They are due to the smooth electrostatic potential evolution that bends electron trajectories away from the lens, as would photons in a gradient index medium.
They almost vanish and match the \emph{loci} of zero charge density in the case of an infinitely sharp circular potential, 
as shown in Fig. \ref{fig4}c for the current density, and concomitantly in Fig. \ref{fig4}d for the calculated SGM map.

\begin{figure}[h]
\includegraphics[width = 1.0 \linewidth]{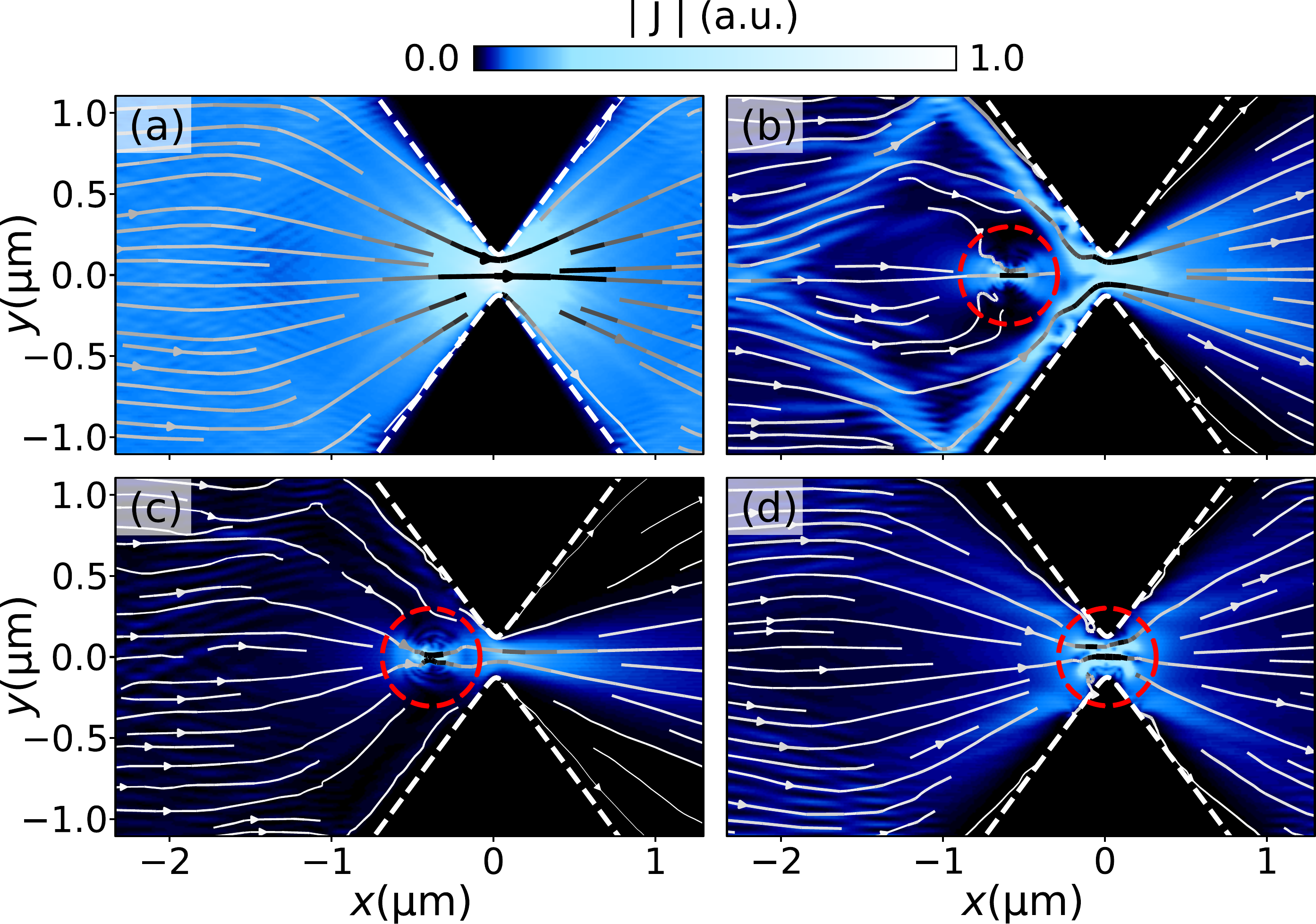}
\caption{\label{fig5} \textbf{Distortion of the flow by the constriction:} (a) Current density through the constriction without tip potential. (b-d) Current density
through the tip and the constriction for different tip-to-constriction distances, indicated in Fig.\ref{fig4}b. The tip potential is the same as in Fig.\ref{fig4}a-b.
}
\end{figure}

Noteworthy, the patterns observed in the SGM maps in the case of a smooth junction are significantly distorted with respect to the corresponding current density patterns. 
For example, the distance from the reduced conductance spot to the constriction center in Fig. \ref{fig4}b is different than the distance from the lowest current density to the lens center in Fig. \ref{fig4}a. 
The spot of enhanced conductance also appears wider than the sharp spot of high current density.
The constriction shape plays a central role in this distortion, as it alters Dirac fermions trajectories.
The finite opening angle of the etched region causes a focusing of current density lines towards the constriction, as depicted in Fig.\ref{fig5}a where we plot the current density $\vec{J}$ through the constriction.
When the tip is over the position of minimum $\arrowvert \vec{J} \arrowvert$  along transport axis (position b in Fig.\ref{fig4}b), the total transmission is still high because of 
trajectories refocused by the constriction edges, as evidenced in Fig.\ref{fig5}b. The lowest total transmission is obtained 
when the tip also blocks these trajectories, as depicted in Fig.\ref{fig5}c. Finally, when placing the tip at the center of the constriction, current density lines are perpendicular to the 
p-n interface (Fig.~\ref{fig5}b) and the efficient Klein tunneling in this configuration leads to a high transmission.
Noticeably, in the case of an abrupt constriction, the simulated SGM images appear more similar to the current density patterns around the Veselago 
lens than in the current geometry (with V-shaped etched regions), which offers promising perspectives for future experiments (See Fig.S9~\cite{Supplemental}).

In conclusion, a SGM tip scanned above a high mobility graphene device is used to create a smooth circular Veselago lens. 
By scanning the tip in the vicinity of a constriction, images related to the electron flow through this circular p-n junction are obtained, as confirmed by tight-binding simulations.  
In particular, a high current density is observed at the lens core, as a direct consequence of Klein tunneling.
The present study also reveals the existence of low current density points away from a smooth circular p-n junction, highly sensitive to the junction smoothness.
While many studies investigated caustics of Veselago lenses \cite{Cserti-2007, Mu-2011,Agrawal-2014, Wu-2014, Logermann-2015, Lu-2018,Zhang-2018} (i.e. points at which trajectories density diverges),
the existence of such low current density points may prove very useful in the future design of Dirac fermions optical elements, adding a new possibility to the existing toolbox.

The present research was funded by the F\'ed\'eration Wallonie-Bruxelles through the ARC Grant on 3D nanoarchitecturing of 2D crystals (No. 16/21-077) and from the European Union's Horizon 2020 Research and Innovation program (No. 696656). B.B. (research assistant), N.M. (FRIA fellowship), B.H. (research associate), V.-H.N. and J.-C.C. (PDR No. T.1077.15) acknowledge financial support from the F.R.S.-FNRS of Belgium.
Support by the Helmholtz Nanoelectronic Facility (HNF), the EU ITN SPINOGRAPH and the DFG (SPP-1459) is gratefully acknowledged. Growth of hexagonal boron nitride crystals was supported by the Elemental Strategy Initiative conducted by the MEXT, Japan and JSPS KAKENHI Grant Numbers JP26248061, JP15K21722 and JP25106006.
B.H. thanks Z. Zanolli for the connection with C.S. B.B aknowledges the use of Kwant \cite{Groth-2014} used to guide the experiment and cross-check tight-binding simulations.

\bibliographystyle{apsrev4-1}

%

\newpage

\onecolumngrid

\includegraphics[page=1,scale=0.85]{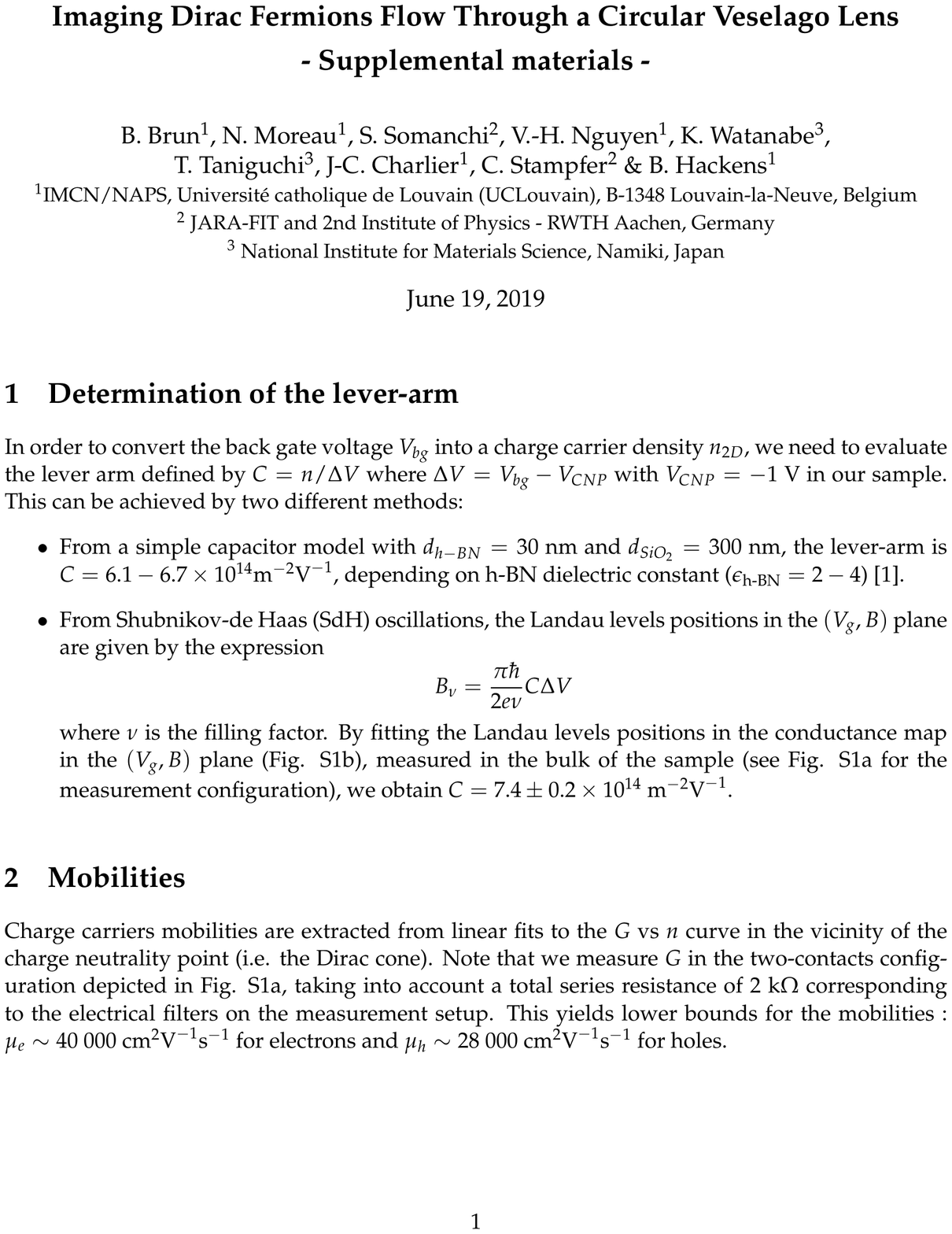} 
\includegraphics[page=2,scale=0.85]{Brun_supplemental.pdf} 
\includegraphics[page=3,scale=0.85]{Brun_supplemental.pdf} 
\includegraphics[page=4,scale=0.85]{Brun_supplemental.pdf} 
\includegraphics[page=5,scale=0.85]{Brun_supplemental.pdf} 
\includegraphics[page=6,scale=0.85]{Brun_supplemental.pdf} 
\includegraphics[page=7,scale=0.85]{Brun_supplemental.pdf} 
\includegraphics[page=8,scale=0.85]{Brun_supplemental.pdf} 
\includegraphics[page=9,scale=0.85]{Brun_supplemental.pdf} 
\includegraphics[page=10,scale=0.85]{Brun_supplemental.pdf} 
\includegraphics[page=11,scale=0.85]{Brun_supplemental.pdf} 
\includegraphics[page=12,scale=0.85]{Brun_supplemental.pdf} 
\includegraphics[page=13,scale=0.85]{Brun_supplemental.pdf} 
\includegraphics[page=14,scale=0.85]{Brun_supplemental.pdf}

\end{document}